\documentclass[11pt,a4paper,english,nofootinbib,,superscriptaddress]{revtex4}
\usepackage{lmodern}

\usepackage[T1]{fontenc}
\usepackage[latin9]{inputenc}
\usepackage{babel}
\usepackage{color}
\usepackage{amsmath}
\usepackage{graphicx}
\usepackage{amssymb}
\usepackage{esint}
\usepackage[unicode=true, pdfusetitle,
bookmarks=true,bookmarksnumbered=false,bookmarksopen=false,
breaklinks=false,pdfborder={0 0 1},backref=false,colorlinks=false]
{hyperref}
\usepackage{amsmath}
\usepackage{amssymb}
\def\barra#1{\not \!#1}
\def\b{\begin{equation}} \def\e{\end{equation}}
\def\bd{\begin{displaystyle}} \def\ed{\end{displaystyle}}
\def\ba{\begin{array}} \def\ea{\end{array}}

\def\bee{\begin{enumerate}}
	\def\eee{\end{enumerate}}

\def\ud{\mathrm{d}}

\def\1{\mbox{I\hspace{-.15em}1}}

\def\R{{\rm I\hspace{-.15em}R}}

\def\b{\begin{equation}}
\def\e{\end{equation}}
\def\bee{\begin{enumerate}}
	\def\eee{\end{enumerate}}

\makeatletter
\usepackage{latexsym}\usepackage{bm}
\usepackage{tikz}
\usetikzlibrary{arrows,decorations.pathmorphing,decorations.markings}

\tikzstyle{midway}=[postaction={decorate},decoration={markings,mark=at position 0.5 with {\arrow{>}}}]

\makeatother

\begin{document}
	\title{Scalar-spinor fields interaction \\ in de Sitter ambient space formalism}
	
	\author{Y. Ahmadi}
	\email{ahmadi.pave@gmail.com} \affiliation{Department of
		Physics, Razi University, Kermanshah, Iran}
	\author{F. Jalilifard}
	\email{jalilifard@iau-ea.ac.ir}\affiliation{Department of Physics, Razi University, Kermanshah, Iran}
	
	\author{M.V. Takook}
	\email{takook@razi.ac.ir} \affiliation{Department of Physics, Razi University, Kermanshah, Iran}
\begin{abstract}

\noindent \hspace{0.35cm}
In de Sitter ambient space formalism, the massless minimally coupled scalar field can be constructed from a massless conformally coupled scalar field and a constant five-vector $A^{\alpha}$. Also, a constant five-vector $B^{\alpha}$ appears in the interaction Lagrangian of massless minimally coupled scalar and spinor fields in this formalism. These constant five-vector fields can be fixed in the interaction case in the null curvature limit. Here we will  calculate the $\cal S$ matrix elements of scalar-spinor fields interaction in the tree level approximation. Then the constant five-vectors $A^{\alpha}$ and $B^{\alpha}$, will be fixed by comparing the $\cal S$ matrix elements in the null curvature limits with the Minkowskian counterparts.
	
\end{abstract}

\maketitle
\section{Introduction}
The interaction Lagrangian in standard model is usually constructed from gauge theory, but the scalar-spinor field interaction is presented by the Yukawa potential. In quantum field theory (QFT) the Yukawa potential cannot be written from any group theoretical point of view. In the previous paper, the interaction Lagrangian of scalar-spinor fields is defined by a new transformation in de Sitter (dS) ambient space formalism, which was free of any infrared divergence \cite{higgs}. The Yukawa potential  can be obtained from this Lagrangian in the null curvature limit. In this model two constant five-vector fields appear, one is from the new transformation and the other is from the massless minimally coupled (\textit{mmc}) scalar field two-point function \cite{higgs}. In this article we will fixed this two constant five-vector fields by calculating the  scattering $(\cal S)$ matrix elements of the scalar-spinor fields interaction in the null curvature limit.

The experimental data confirm that the our universe can be described by the dS metric in the large scale and also in the inflationary epoch \cite{Riess, Perl}.
Thus it is important to construct QFT in dS space-time. In last years, the efforts have been made to do it in ambient space formalism \cite{ta97,77,ta96,taazba,taro12,berotata,derotata}.
The advantage of using ambient space is due to the linearity of the group action on the ambient space formalism, then the calculations are very similar to the Minkowski counterpart.

We know that, in the Minkowski space-time, the analyticity properties of the two-point functions are the fundamental basis for calculation of   the probability amplitude or the Green functions of the interaction fields. The analyticity properties of the two-point functions in dS ambient space formalism have been proved by Bros et al. \cite{brgamo,brmo,brmo03}. 
Therefore, this formalism is more appropriate for constructed the quantum 
interaction fields in dS space-time. In the previous article, we studied the Compton scattering in dS ambient space formalism and showed that the $\cal S$ matrix elements in the null curvature limit reduce to the Minkowskian counterpart \cite{electron-photon}. 
The interaction fields are important in this article for two reason: 
First, the effects of the classical gravitational field on the quantum field  are better seen in the QFT in the dS ambient space formalism. Secondly, in the formulation of the QFT in dS ambient space formalism, two constant five-vector fields appear, which can be fixed in the null curvature limit \cite{ta97,77,higgs}.

The organization of this article is as follow. We present the notation and terminology that has been used in this article in section \ref{notation}. In section \ref{interaction}, the interaction Lagrangian  in the small curvature approximation has been presented and then the $\cal S$ matrix elements for the spinor-scalar interaction have been calculated in dS ambient space formalism. In section \ref{flat limit}, the null curvature limit of the $\cal S$ matrix element has been calculated and then the constant five-vectors $A^\alpha$ and $ B^\alpha$ have been fixed in this limit. Finally, the conclusion has been presented in section \ref{Conclusion}.

\setcounter{equation}{0}	
\section{notation and terminology}\label{notation}

The dS space-time can be considered as a 4-dimensional  hyperboloid embedded in 5-dimensional Minkowski space with the following relation:
$$
M_H=\{x \in \R^5| \; \; x \cdot x=\eta_{\alpha\beta} x^\alpha
x^\beta =-H^{-2}\};\;\; \alpha,\beta=0,1,2,3,4 .$$
The dS metric is:
$$
ds^2=\eta_{\alpha\beta}dx^{\alpha}dx^{\beta}|_{x^2=-H^{-2}}=g_{\mu\nu}^{dS}dX^{\mu}dX^{\nu};\;\; \mu=0,1,2,3 ,$$
where
 $\eta_{\alpha\beta}=$diag$(1,-1,-1,-1,-1)$, $H$ is Hubble parameter, $X^\mu$ is dS intrinsic coordinate and $x^\alpha$ is the 5-dimensional dS ambient space formalism ($x\cdot x=-H^{-2}$). The ambient space coordinate $x^\alpha$ can be represented in terms of the intrinsic coordinate $X^\mu$. For defining time evolution operator and scattering matrix, one can choose the following static coordinate system:
 \b \label{coordinate}
 \left\{\begin{array}{clcr} x^0&=\sqrt{H^{-2}-r^2}\sinh Ht_s, \\                    
 	x^1&=\sqrt{H^{-2}-r^2}\cosh Ht_s, \\
 	x^2&=r\cos \theta , \\
 	x^3&=r \sin \theta\cos\phi ,\\   
 	x^4&=r\sin\theta\sin \phi,\\            
 \end{array} \right.\e
 where $-\infty<t_s<\infty\;,\;0\leq r<H^{-1}\;,\;0\leq \theta\leq\pi\;,\;0\leq \phi< 2\pi$.  This coordinate system does not covered  the total of the dS hyperboloid. For investigating the null curvature limit, the following coordinate system is useful:
  \b \label{II.7}
 \left\{\begin{array}{clcr} x^0&=H^{-1}\sinh(HX^0), \\                    
 	\vec x&=H^{-1}\dfrac{\overrightarrow{X}}{\lVert\overrightarrow{X}\lVert}\cosh(HX^0)\sinh(H\lVert\overrightarrow{X}\lVert), \\
 	x^4&=H^{-1}\cosh(HX^0)\cosh(H\lVert\overrightarrow{X}\lVert),\\            
 \end{array} \right.\e
where the $\lVert\overrightarrow{X}\lVert=(X_1^2+X_2^2+X_3^3)^\frac{1}{2}$ is the norm of three-vector $\overrightarrow{X}$.

 The Fourier transformation cannot be generally defined in curved space-time, but in the  dS space-time, due to the maximally symmetric properties of dS hyperboloid the Fourier-Helgason-type transformation or the Bros-Fourier transformation can be defined \cite{hel62,hel94,brmo03,brmo}. Then, corresponding to any space-time variable $x^\alpha$ another transformed variable, $\xi^\alpha=(\xi^0, \vec \xi, \xi^4)$ can be defined in the positive null-cone $C^+=\left\lbrace \xi \in \R^5|\;\; \xi\cdot \xi=0,\;\; \xi^{0}>0 \right\rbrace$ \cite{brmo03,brmo}. One can parameterize the  $\xi$ in terms of massive particle four-momentum in flat Minkowski space-time as:
\b \label{xi expression}\xi=\left(\xi^{0} , \overrightarrow{\xi} , \xi^{4}\right)=\left(\dfrac{k^{0}}{m}=\sqrt{\dfrac{\overrightarrow{k}^2}{m^2}+1} , \dfrac{\overrightarrow{k}}{m} , -1\right).\e

In the dS ambient space formalism, dS-Dirac first-order field equation is \cite{bagamota}:
\b \label{ds dirac equation}
\left(-i\barra{x} \barra{\partial}^\top+2i \pm \nu\right)\psi(x)=0,\e
where $\barra x=\eta_{\alpha\beta}\gamma^{\alpha} x^{\beta}$ and $\partial^{\top}_{\alpha}=\theta_{\alpha \beta}\partial^{\beta}=\partial_{\alpha}+Hx_{\alpha}x\cdot\partial$. $\theta_{\alpha \beta}=\eta_{\alpha\beta}+H^2x_{\alpha}x_{\beta}$ is the projection operator on dS hyperboloid. In this equation, $\nu$ is related to dS mass parameter $m^2_{f,\nu}=H^2(2+\nu^2\pm i\nu)$ and in the null curvature limit it is defined the Minkowskian mass parameter $m$ \cite{77}:
$$\lim_{H\rightarrow 0,\nu\rightarrow \infty} (H\nu)^2=m^2. $$ The five matrices $\gamma^{\alpha}$, which satisfy the conditions 
$
\gamma^{\alpha}\gamma^{\beta}+\gamma^{\beta}\gamma^{\alpha}
=2\eta^{\alpha\beta}$ and $\gamma^{\alpha\dagger}=\gamma^{0}\gamma^{\alpha}\gamma^{0}$, can be chosen as \cite{ta97,ta96,bagamota}:
$$ \gamma^0=\left( \begin{array}{clcr} I & \;\;0 \\ 0 &-I \\ \end{array} \right)
,\;\;\;\gamma^4=\left( \begin{array}{clcr} 0 & I \\ -I &0 \\ \end{array} \right) , $$
\b \label{II.9}
\gamma^1=\left( \begin{array}{clcr} 0 & i\sigma^1 \\ i\sigma^1 &0 \\
\end{array} \right)
,\;\;\gamma^2=\left( \begin{array}{clcr} 0 & -i\sigma^2 \\ -i\sigma^2 &0 \\
\end{array} \right)
, \;\;\gamma^3=\left( \begin{array}{clcr} 0 & i\sigma^3 \\ i\sigma^3 &0 \\
\end{array} \right),\e
where $\sigma^i$ $(i=1,2,3)$ are the Pauli matrices. These matrices are different from Minkowski gamma matrices $\gamma'^{\mu}$. The relation between the $\gamma$ matrices in Minkowski and dS ambient space formalism are \cite{bagamota}:
\b \label{gamma relation}\gamma'^{\mu}=\gamma^{\mu}\gamma^4.\e

The charged spinor field operator, which satisfies the equation (\ref{ds dirac equation}),  is \cite{77}:
\b \label{psi expansion}\psi(x)=\int_{S^3}d\mu(\xi)\sum_{n}\left[a( \tilde{\xi},n)(Hx.\xi)^{-2-i\nu} {\cal U} (x,\xi,n)+b^{\dag}(\xi,n)(Hx.\xi)^{-1+i\nu} {\cal V} (x,\xi,n)\right],\e
where $ n=-\frac{1}{2},\frac{1}{2}$, $\ud\mu(\xi)$ is the $SO(4)$-invariant normalized volume element and $\tilde{\xi}^\alpha=(\xi^0, -\vec \xi, \xi^4)$. The explicit form of $\cal{U}$ and $\cal{V}$ has been defined in \cite{ta97,bagamota}. The adjoint spinor $\overline{\psi}(x)$ in ambient space formalism is defined as $\overline{\psi}(x)=\psi^\dagger(x)\gamma^0\gamma^4$ \cite{bagamota}. The $a( \tilde{\xi},n),\; a^{\dagger}( \tilde{\xi},n)\; \text{and} \; b( \tilde{\xi},n),\; b^{\dagger}( \tilde{\xi},n)$ are annihilation and creation operators \cite{77}:
\b \label{creation operator}
a^\dag (\xi,n) \left| \Omega \right> \equiv\left|1_{\xi, n}^{a} \right\rangle,\;\; b^\dag (\xi,n) \left| \Omega \right> \equiv\left|1_{\xi, n}^{b} \right\rangle\, .\e
The vacuum state $|\Omega>$, with norm  $\left< \Omega \right.\left| \Omega \right>=1 $, is invariant under the action of the UIR of the dS group. This vacuum state can be identified with the Bunch-Davies or Hawking-Ellis vacuum state. The anti-commutation relations for creation and annihilation operators are \cite{bagamota,77}:
$$\left[a(\tilde{\xi}',n') , a^\dagger(\xi,n)\right]_{+}={\cal N}_p(\xi,n)\delta_{s^3}(\xi-\xi')\delta_{nn'},$$
\b \label{anti-commutation relation}
\left[b(\tilde{\xi}',n') , b^\dagger(\xi,n)\right]_{+}={\cal N}_q(\xi,n)\delta_{s^3}(\xi-\xi')\delta_{nn'}, \e
where ${\cal N} _p(\xi,r)\equiv{\cal N} _p$ and ${\cal N} _q(\xi,r)\equiv{\cal N} _q$ are the normalization constants.

In dS ambient space formalism, the field equation for massless conformally coupled (\textit{mcc}) scalar field is \cite{brmo,ta97,77}:
\b \label{II.37}
\left(Q_0-2\right) \phi_{mcc}(x)=0 ,\e
that $Q_0=-H^{-2}\partial^\top\cdot\partial^\top \equiv -H^{-2} \square_H\;.$ The \textit{mcc} scalar field operator, which satisfies the field equation (\ref{II.37}), is \cite{brmo,77}:
\b \label{mcc expansion}
\phi_{mcc}(x)=\sqrt{ c_0 }\int_{S^3}d\mu({\xi})\left[d({ \tilde{\xi}},1)(Hx\cdot\xi)^{-2}+d^\dag(\xi,0)(Hx\cdot\xi)^{-1}\right],\e
where $c_0$ is the normalization constant. The $d(\tilde{\xi},1)$ and $d^\dag (\xi,0)$  are annihilation and creation operators respectively:
$$d^\dag (\xi,0) \left| \Omega \right> \equiv\left|1_{\xi}^{d} \right\rangle,\;\;d(\tilde{\xi},1) \left| \Omega \right>=0.$$
Between the massless minimally coupled (\textit{mmc}) scalar field, and the \textit{mcc} scalar fields in the dS ambient space formalism, the following relation exist \cite{higgs,77,khrota,gareta}:
\b\label{magic}
\phi_{mmc}(x)= \left[A\cdot\partial^\top + 2 A\cdot H^2x\right]\phi_{mcc}(x),\e
where $A^\alpha$ is an arbitrary constant five-vector.
The \textit{mmc} scalar field operator can be obtained by inserting (\ref{mcc expansion}) in (\ref{magic}) and doing some simple calculations \cite{higgs,77}. The analytic field operator is defined in complex dS space-time $\phi_{mmc}(x)=\lim_{y\rightarrow 0}\phi_{mmc}(z)=\lim_{y\rightarrow 0}\phi_{mmc}(x+iy).$ 

The analytic two-point function of \textit{mcc} scalar field can be calculated in terms of the generalized Legendre function \cite{brmo,brgamo}:
$$W_{mcc}(z,z')=\left<\Omega|\phi_{mcc}(z)\phi_{mcc}(z')|\Omega\right>=$$
\b \label{mcc 2point function}
c_0\int d\mu(\xi) (Hz\cdot\xi)^{-2}(Hz'.\cdot\xi)^{-1}=\dfrac{-H^2}{8\pi^2}\dfrac{1}{1-{\cal Z}(z,z')}=\dfrac{1}{(2\pi)^2}\dfrac{1}{(z-z')^2}\, ,\e
where ${\cal Z}(z,z')=1+\dfrac{H^2}{2}(z-z')^2$ is the geodesic distance between two point $z$ and $z'$  \cite{brmo,chta}.

One can define the \textit{mmc} scalar analytic two-point function in terms of \textit{mcc} scalar analytic two-point function as \cite{77,higgs}:
\b \label{mmc 2point function} W_{mmc}(z,z') =\left[A\cdot\partial^\top + 2 A\cdot H^2z\right]\left[A\cdot\partial'^\top + 2 A\cdot H^2z'\right] W_{mcc}(z,z').\e  
For defining the real two-point function one can use the boundary value of the analytic two-point function $ W_{mmc}(z,z')$ \cite{77,khrota},
$${\cal W}_{mmc}(x,x')=\lim_{y,y'\rightarrow0} W_{mmc}(z,z')=\lim_{y,y'\rightarrow0} W_{mmc}(x+iy\;,\;x'+iy').$$

\section{scattering matrix}\label{interaction}

In ambient space formalism, the scalar field equation in dS space-time can be written as: $$(Q_0+\sigma(\sigma+3))\phi(x)=0\,. $$ The solutions of this equation are named dS plane waves $(x\cdot\xi)^\sigma$. The five-vector $\xi^\alpha$ plays the role of momentum parameter and it is lying in the null-cone in ambient space \cite{higgs}. The action of {\it mcc} scalar and spinor free fields in dS universe is \cite{77,bagamota,higgs}:
\b \label{conf spinor massless action} S(\psi,\phi)=\int \ud\mu(x){\cal L}(\psi , \phi)=\int \ud\mu(x)\left[H \overline{\psi} \gamma^4\left( -i\barra{x} \barra{\partial}^\top+2i\right) \psi+\phi \;Q_0\;\phi \right],\e where $ \ud\mu(x)$ is dS invariant volume element. The above action is invariant under the global U(1) symmetry. By defining the covariant derivative $D_\alpha\psi=(\partial_\alpha^\top+i{\cal G}B^\top_\alpha\phi_{mmc})\psi $, where $B^\alpha B_\alpha=0$ and ${\cal G}$ is the coupling constant that determines interaction intensity, the action (\ref{conf spinor massless action}) will be invariant under the following transformations \cite{higgs}:
$$\left\{\begin{array}{clcr}
	\phi \rightarrow \phi'=\phi+(x.B)^{-3} \\
	\psi \rightarrow \psi'=e^{\frac{i}{2} {\cal G}(x\cdot B) ^{-2}}\psi\;\;\;\;\;\;\;\;\;       
\end{array} \right. .$$
So by replacing this new derivative $D_\alpha$ with transverse derivative $\partial_\alpha^\top$, the interaction Lagrangian is obtained as \cite{higgs}:
 \b \label{higgs spinor lagrangian} {\cal L}_{int}={\cal G}\;H\overline{\psi} \gamma^4 \barra{x} \barra{B^\top}\phi_{mmc}\psi. \e 
The constant five-vector $B^\alpha$ in Lagrangian and constant five-vector $A^\alpha$  in {\it mmc} scalar two-point function can be fixed in the interaction case in the null curvature limit. The interaction Lagrangian plays an important role for calculation of the probability amplitude or the $\cal S$ matrix elements.

The time evolution operator cannot be explicitly defined in a general curved space time. In previous article we show that the time evolution operator $U(t,t_0),$ such that $\vert \alpha , t\rangle=U(t,t_0) \vert \alpha ,t_0\rangle$, can be defined in the dS static coordinate system  (\ref{coordinate}), but it is very complicated for calculation of the $\cal S$ matrix element \cite{bh}:
\b \label{smatrix} |\phi_{out}\;,+\infty>=U(\infty\;,-\infty)|\phi_{in}\;,-\infty>= {\cal S}|\phi_{in}\;,-\infty>.\e 

Then we presented a logical approximation when the interaction occurs in the atomic dimension \cite{electron-photon}, which the $\cal S$ matrix elements can be calculated. In this approximation the direct effect of curvature can be ignored but the indirect effect is presented. In this case the time evolution operator can be expanded on the Minkowski space-time as \cite{electron-photon}:
$$U(t,t_0)={U_M}(t,t_0)+ Hf(t,t_0)+\cdots\;,$$
 the term $f(t,t_0)$ is due to the direct effect of curvature that in atomic dimension and in  null curvature limit, it is disappeared. The  mathematical form of $U_M$ is exactly similar to the time evolution operator in Minkowskian space-time. Therefore the scattering matrix may be presented with the following approximate equation from the Minkowskian counterpart
\cite{mandl shaw, kaku,zuber}:
$${\cal S}=\frac{(-1)^{\tau}}{{\tau}!} \sum_{{\tau}=0}^{\infty}{\cal S}^{({\tau})}+\cdots\, , $$
where
\b \label{s matrix expansion} {\cal S}^{({\tau})}=\int\ud\mu(x_1)\int\ud\mu(x_2)\cdots \int\ud\mu(x_{\tau})\;\;\; T[{\cal L}_{int}(\phi(x_1)){\cal L}_{int}(\phi(x_2))\cdots {\cal L}_{int}(\phi(x_{\tau}))] .\e
In this equation, ${\cal L}_{int}(\phi(x))$ is the interaction Lagrangian and $T[{\cal L}_{int}(\phi(x_1)){\cal L}_{int}(\phi(x_2)){\cal L}_{int}(\phi(x_3))...]$ is the time order product of fields.


By inserting interaction Lagrangian of spinor-scalar fields (\ref{higgs spinor lagrangian}) in (\ref{s matrix expansion}), the ${\cal S}^{(2)}$ is obtained as:
\b \label{s2 time order}
{\cal S}^{(2)}={\cal G}^2H^2\int\int\ud\mu(x) \ud\mu(x')\;\;T\left[\left(\overline{\psi}\gamma^4\barra{x}\barra{B}^\top\phi_{mmc}\psi\right)_{x}\left(\overline{\psi}\gamma^4\barra{x'}\barra{B}^\top\phi_{mmc}\psi\right)_{x'}\right].\e

By using Wick's theorem, one can write the time order product of fields in terms of the normal order product of fields. After some simple calculations, the time order product of (\ref{s2 time order}) turns into 8 normal order product. Since we consider the interaction, which is depicted in figure \ref{f1}, only one term is remained :

\begin{figure}[!ht]
	
	\begin{tikzpicture}[scale=1]
	\draw[midway,very thick,blue](0,0) -- (-1.3,1.3);
	\draw[midway,very thick,red](1.5,0) -- (2.8,1.3);
	\draw[midway,very thick,red] (-1.3,-1.3)--(0,0);
	\draw[->,dashed,very thick] (0,0)--(0.75,0);
	\draw[dashed,very thick](0.8,0)--(1.5,0);
	\draw[midway,very thick,blue] (2.8,-1.3)--(1.5,0);
	\node (a1) at (-1.3,0.8) {$q$};
	\node (a2) at (-1.3,-0.8) {$p$};
	\node (a3) at (0.75,-0.78) {$p+q$};
	\node (a4) at (2.8,0.8) {$p'$};
	\node (a5) at (2.8,-0.8) {$q'$};
	\node (a7) at (-0.4,0) {$x$};
	\node (a8) at (1.9,0) {$x'$};
	\end{tikzpicture}
	\caption {spinor-spinor interaction diagram with {\it mmc} scalar propagator}
	\label{f1}
\end{figure}
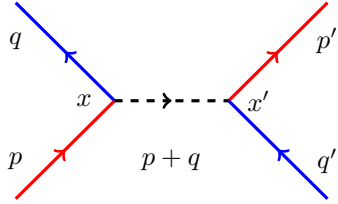
\b\label{s2}
{\cal S}^{(2)}={\cal G}^2H^2\int \int \ud\mu(x) \ud\mu(x')\;\;N\left[(\overline{\psi}\gamma^4\barra{x}\barra{B}^\top\underleftrightarrow{\phi_{mmc}\psi)_{x}(\overline{\psi}\gamma^4\barra{x'}\barra{B}^\top\phi}_{mmc}\psi)_{x'}\right],\e
where, the symbol $\leftrightarrow$ is the \textit{mmc} scalar two-point function (\ref{mmc 2point function}) and $N[...]$ is normal order product,
$$\underleftrightarrow{\phi_{mmc}(x)\;\;\phi}_{mmc}(x')=\left<\Omega|\phi_{mmc}(x)\;\;\phi_{mmc}(x')|\Omega\right>={\cal W}_{mmc}(x,x').$$
Since in (\ref{s2}) the spinor field, $\gamma$ matrices and other quantities are matrix, one can write them in terms of their components:
$$ {\cal S}^{(2)}={\cal G}^2 H^2 \int\int\ud\mu(x)\ud\mu(x')\;{\cal W}_{mmc}(x , x')\left(\gamma^4\barra{x}\barra{B^\top}(x)\right)_{gh}\left(\gamma^4\barra{x'}\barra{B^\top}(x')\right)_{kl}\;$$
\b \label{normal order product}\times N\left[\overline{\psi}_g(x)\psi_h(x)\overline{\psi}_k(x')\psi_l(x')\right].
\e
By breaking the spinor field (\ref{psi expansion}) into $+, -$ parts, the normal order product in above relation (\ref{normal order product}) converted to 16 terms but just one of them is described the figure \ref{f1},
\b \label{+,-}\psi(x)=\psi^+(x)+\psi^-(x)
,\;\;\;\;\;
\overline{\psi}(x)=\overline{\psi}^+(x)+\overline{\psi}^-(x)
\e
$$\mbox{physical term in normal order product}=\overline{\psi}^-_g(x)\psi^+_h(x)\psi^-_l(x')\overline{\psi}^+_k(x').$$
Thus the ${\cal S}^{(2)}$ is obtained as:
$$ {\cal S}^{(2)}={\cal G}^2 H^2 \int\int\ud\mu(x)\ud\mu(x')\;{\cal W}_{mmc}(x , x')\;\left(\gamma^4\barra{x}\barra{B^\top}(x)\right)_{gh}\left(\gamma^4\barra{x'}\barra{B^\top}(x')\right)_{kl}$$
$$\times\{\overline{\psi}^-_g(x)\psi^+_h(x)\psi^-_l(x')\overline{\psi}^+_k(x')\}.
$$

In this interaction, incoming sate
$|i>$ is two spinor fields with $\xi_p , \xi_q$ ''momentums'' and $r, n$spin polarizations. Also outgoing state $|f>$ is two spinor fields with  $\xi'_p, \xi'_q$ ''momentums'' and $r', n'$ spin polarizations,
$$|i>=|1_{p}^{a^{\dagger}} , 1_{q}^{b^{\dagger}}>=a^{\dagger}_r(\xi_{p})b^{\dagger}_n(\xi_{q})|\Omega>$$ 
$$|f>=|1_{p'}^{a^{\dagger}} , 1_{q'}^{b^{\dagger}}>=a^{\dagger}_{r'}({\xi}'_{p})b^{\dagger}_{n'}({\xi}'_{q})|\Omega>\;\;\;\;\; \Longrightarrow \;\;\;\;\; <f|=<1_{p'}^{a} , 1_{q'}^{b}|=<\Omega|b_{n'}({\xi}'_{q})a_{r'}({\xi}'_{p}).$$
Then the ${\cal S}^{(2)}_{fi}$ is:
$$ {\cal S}^{(2)}_{fi}=<f|{\cal S}^{(2)}|i>={\cal N}_{p}{\cal N}_{p'}{\cal N}_{q}{\cal N}_{q'} {\cal G}^2 H^2 $$
$$\times\int\int\ud\mu(x)\;\ud\mu(x')\;{\cal W}_{mmc}(x , x')\;\left(\gamma^4\barra{x}\barra{B^\top}(x)\right)_{gh}\left(\gamma^4\barra{x'}\barra{B^\top}(x')\right)_{kl}$$
$$\times(Hx.{\xi}_{q})^{-1-i\nu} (Hx.\xi_{p})^{-2-i\nu}(Hx'.\xi'_{q})^{-1+i\nu} (Hx'.\xi'_{p})^{-2+i\nu}$$
 \b \label{s matrix} \times \overline{\cal V}_g(x , \xi_q , n)\;\;{\cal U}_h(x , \xi_{p} , r)\;\;\overline{{\cal U}}_k(x' , \xi'_{p} , r')\;\;{\cal V}_l(x' , \xi'_{q} , n').\e
 
 This $\cal S$ matrix elements may be calculated by numerical methods, but in this paper we would like to fix the constant five-vectors $A^\alpha$ and $B^\alpha$, which will be discussed in the next section.

\section{flat limit}\label{flat limit}

The radius of dS hyperboloid is $H^{-1}$. The curvature of space-time diminishes for large radius, or equivalently ${H\rightarrow0}$, and dS space-time matches with Minkowski space-time. In this limit one can obtain these relations \cite{ta97}:
$$\lim_{H\rightarrow 0}(x\cdot\xi)^{-2-i\nu}=e^{-ik.X} ,\;\;\;\;\;\;\;\;\;\;\lim_{H\rightarrow 0}H\barra{x}=\lim_{H\rightarrow 0}H\eta_{\alpha\beta}\gamma^\alpha x^\beta=-\gamma^4,$$
$$\lim_{H\rightarrow 0}\theta_{\alpha\beta}=\lim_{H\rightarrow 0}\left(\eta_{\alpha\beta}+H^2x_{\alpha}x_{\beta}\right)=\eta_{\mu\nu}, \;\;\;\;\;\; \left\{ \begin{array}{clcr} \mu=\nu=0,1,2,3\;\;\;
\\
\alpha=\beta=0,1,2,3,4\;.\\ \end{array} \right.$$
Since the five-vector $B^\alpha$ is constant, at null curvature limit $B^4$ can be chosen as zero and then
$$\lim_{H\rightarrow 0}\barra B^{\top}=-B_{\mu}\gamma^{'\mu}\gamma^4\;;\;\;\;\;\;\mu=0,1,2,3.$$
The Minkowski $\gamma'$ matrices are related to ambient $\gamma$ matrices as (\ref{gamma relation}).
The analytic \textit{mmc} scalar two-point function (\ref{mmc 2point function}) in null curvature limit, $H\rightarrow 0$, is:
$$\lim_{H\rightarrow 0}W_{mmc}(z,z') =\lim_{H\rightarrow 0}\left[A\cdot\partial^\top + 2 A\cdot H^2z\right]\left[A\cdot\partial'^\top + 2 A\cdot H^2z'\right] W_{mcc}(z,z')$$ $$=(A\cdot\partial) (A\cdot\partial') \lim_{H\rightarrow 0} W_{mcc}(z,z')
$$
and by using the (\ref{mcc 2point function}), the real \textit{mmc} scalar two-point function in null curvature limit, $H\rightarrow 0$, obtains as:
$$\lim_{H\longrightarrow 0}{\cal W}_{mmc}(x,x')\equiv w(X,X')=\dfrac{1}{(2\pi)^2}(A.\partial)(A.\partial')\dfrac{1}{(X-X')^2}\;.$$

By using this result for \textit{mmc} scalar field, adjoint spinor definition in ambient space formalism (\ref{psi expansion}) and delta function definition in Minkowski space-time , $\delta^4(p-p')=\int\dfrac{\ud^4X}{(2{\pi})^4}e^{-iX.(p-p')}$, the the null curvature limit of ${\cal S}^{(2)}_{fi}$ obtains as:
$$ \lim_{H\rightarrow 0}{\cal S}^{(2)}_{fi}=\dfrac{1}{(2\pi)^2}{\cal N}_{p}{\cal N}_{p'}{\cal N}_{q}{\cal N}_{q'}{\cal G}^2
\left\lbrace\int\int\ud^4X\;\ud^4X'e^{-iX.(p+q)}
e^{iX'.(p'+q')}\;\;\; (A.\partial)(A.\partial')\dfrac{1}{(X-X')^2}\right\rbrace$$
\b \label{lim s}\times\left({\cal V}^{\dagger}(q , n)\gamma'^0\right)_{g}\;B_{\mu}\left(\gamma'^{\mu}\gamma^4\right)_{gh}\;{\cal U}_h(p , r)\;\;\;\left({\cal U}^{\dagger}(p', r')\gamma'^0\right)_{k}B_{\nu}\left(\gamma'^{\nu}\gamma^4\right)_{kl}\;{\cal V}_l(q' , n')
.\e
By notice that $\int d^4 X\partial_{\mu}\left(e^{-iX.(p+q)}\dfrac{1}{(X-X')^2}\right)=0$, one can prove the following relation:
$$\int d^4X e^{-iX.(p+q)}\;\partial_{\mu}\left(\dfrac{1}{(X-X')^2}\right)=-\int d^4X \dfrac{1}{(X-X')^2}\left(\partial_{\mu}e^{-iX.(p+q)}\right)$$ $$=-i(p+q)_{\mu}\int d^4X \dfrac{e^{-iX.(p+q)}}{(X-X')^2},$$
then we obtain:
$$\int\int\ud^4X\;\ud^4X'e^{-iX.(p+q)}
e^{iX'.(p'+q')}\;\;\; (A.\partial)(A.\partial')\dfrac{1}{(X-X')^2}=$$
$$A.(p+q)A.(p'+q')\int\int\ud^4X\;\ud^4X'e^{-iX.(p+q)}
e^{iX'.(p'+q')}\;\;\;\dfrac{1}{(X-X')^2}.$$
After some straightforward calculations one can arrive the following relation:
$$\int\int\ud^4X\;\ud^4X'e^{-iX.(p+q)}
e^{iX'.(p'+q')}\;\;\;\dfrac{1}{(X-X')^2}=(2\pi)^8 \delta(p+q-p'-q') \dfrac{1}{(p+q)^2+i\epsilon}.$$
By inserting these results in (\ref{lim s}) one can obtain the null curvature limit of ${\cal S}^{(2)}$ matrix as follow:
$$ \lim_{H\rightarrow 0}{\cal S}^{(2)}_{fi}=(2\pi)^6{\cal N}_{p}{\cal N}_{p'}{\cal N}_{q}{\cal N}_{q'}{\cal G}^2 \delta(p+q-p'-q')$$
$$\times\left({\cal V}^{\dagger}(q , n)\gamma'^0\right)_{g}\;B_{\mu}\left(\gamma'^{\mu}\gamma^4\right)_{gh}\;{\cal U}_h(p , r)\;\;\;\dfrac{[A.(p+q)]^2}{(p+q)^2+i\epsilon}\;\;\;\left({\cal U}^{\dagger}(p', r')\gamma'^0\right)_{k}B_{\nu}\left(\gamma'^{\nu}\gamma^4\right)_{kl}\;{\cal V}_l(q' , n').
$$
By comparing this equation with the Minkowski result \cite{mandl shaw}, the constant five-vectors $A_\mu$ and $B_\mu$ can be fixed as:
$$A_{\mu}=a\;\dfrac{(p_{\mu}+q_{\mu})}{(p+q).(p+q)} \, , \;\;\;\;\; B_{\mu}=b\;\gamma'_{\mu}\gamma^4,$$
where $a$ and $b$ can be written in therms of the normalization constants.

\section{Conclusion} \label{Conclusion}

A constant five-vector field $A^\alpha$ appears in dS \textit{mmc} scalar field quantization. Also another constant five-vector field $B^\alpha$ appears in definition of the interaction Lagrangian of dS spinor-scalar fields interaction. These  constant five-vectors $A^\alpha$ and $B^\alpha$ can be fixed in calculation of the ${\cal S }$ matrix elements in tree diagrams approximation in the null curvature limit. The constant five-vector $B^\alpha$ come from the gauge theory and depended on $\gamma$ matrices and the constant five-vector $A^\alpha$ come from the propagator. It is interesting to note that the constant five-vector $A^\alpha$ depended on the laboratory energy scale i.e. the energy-momentum of the incoming fields, which is a logical result.
These calculations may be open the door to measuring the gravitational effects on quantum fields which may be measurable in the laboratory scale of energy. The other diagrams and interactions may be investigated in the next works.

{\bf{Acknowledgments}}:  The authors wish to express their particular thanks to  M. Amiri, M. Rastiveis,  R. Raziani and S. Tehrani-Nasab for discussions.

\end{document}